\documentclass[sigconf, nonacm, pdfa]{acmart}

\usepackage[a-2b]{pdfx}
\usepackage{algorithmic}
\usepackage{lipsum}
\usepackage{cleveref}
\usepackage{xspace}
\usepackage{listings}
\usepackage{balance}
\usepackage{ragged2e}
\usepackage{tabularx}
\usepackage{svg}

\newcommand\vldbdoi{10.14778/3685800.3685855}
\newcommand\vldbpages{4281 - 4284}
\newcommand\vldbvolume{17}
\newcommand\vldbissue{12}
\newcommand\vldbyear{2024}
\newcommand\vldbauthors{\authors}
\newcommand\vldbtitle{\shorttitle} 
\newcommand\vldbavailabilityurl{https://github.com/DeanLight/spannerlib}
\newcommand\vldbpagestyle{empty} 

\begin{document}

\title{SpannerLib: Embedding Declarative Information Extraction in an Imperative Workflow}
\author{Dean Light}
\affiliation{%
  \institution{Technion}
  \city{Haifa}
  \state{Israel}
}
\email{dean.light92@gmail.com}

\author{Ahmad Aiashy}
\affiliation{%
  \institution{Technion}
  \city{Haifa}
  \state{Israel}
}
\email{ahmad-ai@campus.technion.ac.il}

\author{Mahmoud Diab}
\affiliation{%
  \institution{Technion}
  \city{Haifa}
  \state{Israel}
}
\email{mahmoud.diab@campus.technion.ac.il}

\author{Daniel Nachmias}
\affiliation{%
  \institution{Technion}
  \city{Haifa}
  \state{Israel}
}
\email{nach.daniel@gmail.com}

\author{Stijn Vansummeren}
\affiliation{%
  \institution{UHasselt, Data Science Institute}
  \city{Diepenbeek}
  \state{Belgium}
}
\email{stijn.vansummeren@uhasselt.be }

\author{Benny Kimelfeld}
\affiliation{%
  \institution{Technion}
  \city{Haifa}
  \state{Israel}
}
\email{bennyk@cs.technion.ac.il}

\def\libname{\textsc{SpannerLib}\xspace}

\def\benny#1{{\color{blue}BK: #1}}
\def\dean#1{{\color{red}DL: #1}}
\def\stijn#1{{\color{brown}SV: #1}}

\begin{abstract}
Document spanners have been proposed as a formal framework for declarative Information Extraction (IE) from text, following IE products from the industry and academia. 
Over the past decade, the framework has been studied thoroughly in terms of expressive power, complexity, and the ability to naturally combine text analysis with relational querying. This demonstration presents \libname---a library for embedding document spanners in Python code. \libname facilitates the development of IE programs by providing an implementation of Spannerlog (Datalog-based document spanners) that interacts with the Python code in two directions: rules can be embedded inside Python, and they can invoke custom Python code (e.g., calls to ML-based NLP models) via user-defined functions.
The demonstration scenarios showcase IE programs, with increasing levels of complexity, within Jupyter Notebook.



\end{abstract}

\maketitle

\pagestyle{\vldbpagestyle}
\begingroup\small\noindent\raggedright\textbf{PVLDB Reference Format:}\\
\vldbauthors. \vldbtitle. PVLDB, \vldbvolume(\vldbissue): \vldbpages, \vldbyear.\\
\href{https://doi.org/\vldbdoi}{doi:\vldbdoi}
\endgroup

\begingroup
\renewcommand\thefootnote{}\footnote{\noindent
This work is licensed under the Creative Commons BY-NC-ND 4.0 International License. Visit \url{https://creativecommons.org/licenses/by-nc-nd/4.0/} to view a copy of this license. For any use beyond those covered by this license, obtain permission by emailing \href{mailto:info@vldb.org}{info@vldb.org}. Copyright is held by the owner/author(s). Publication rights licensed to the VLDB Endowment. \\
\raggedright Proceedings of the VLDB Endowment, Vol. \vldbvolume, No. \vldbissue\ %
ISSN 2150-8097. \\
\href{https://doi.org/\vldbdoi}{doi:\vldbdoi} \\
}\addtocounter{footnote}{-1}\endgroup

\ifdefempty{\vldbavailabilityurl}{}{
\vspace{.3cm}
\begingroup\small\noindent\raggedright\textbf{PVLDB Artifact Availability:}\\
The source code, data, and/or other artifacts have been made available at \url{\vldbavailabilityurl}.
\endgroup
}

\section{Introduction}

Historically, approaches to Information Extraction (IE) have been divided into two paradigms: \emph{rule-based} IE, such as IBM's SystemT~\cite{chiticariu-etal-2018-systemt}, and \emph{ML-based} IE of which Transformers and LLMs are the predominant example. While the latter have shown impressive capabilities, it is increasingly recognized that to reach acceptable accuracy in practical scenarios one has to orchestrate LLM-based extractors using scenario-specific logic. This logic is often implemented imperatively in languages such as python. In this paper, we present SpannerLib, a rule-based approach to IE that facilitates the orchestration of ML-based IE in a declarative framework.

\emph{Document spanners} have been studied over the past decade as the theoretical core of industrial systems for Information Extraction (IE) from text~\cite{faginformal} such as IBM's SystemT. The framework casts IE as relational querying: generic IE functions (such as regular expressions with capture variables) extract base relations that, in turn, are manipulated by the relational algebra. Thus, the framework allows to utilize the simplicity and popularity of query languages, such SQL, to greatly simplify the development of solutions for text analysis. The logical nature of document spanners does not preclude the potential of utilizing generic solutions for Natural Language Processing (NLP) that are nowadays dominated by statistical and numerical analysis, from basic classifiers to Large Language Models (LLMs). Extensions such as Spannerlog~\cite{nahshonIncorporatingInformationExtraction2016} showed how the theoretical framework can be elegantly combined with custom code as special user-defined functions called \emph{IE functions}. 

In this demonstration, we showcase \libname: a library for integrating document spanners in Python, resulting in an IE development that enjoys the benefits of the \emph{imperative} paradigm (arbitrary Python code, including calling ML models) and the \emph{declarative} one (Datalog), thereby making rule-based IE development accessible to a wider developer audience. We describe three pillars of our system: \emph{(1)} a full-fledged implementation of Spannerlog, \emph{(2)}, an extension of Python with embedded Spannerlog via iPython's Magic system, where the two types of code communicate via Pandas DataFrames, \emph{and (3)} a framework for constructing and registering  IE functions to invoke Python code from within Spannerlog rules, which now serve as \emph{callback} functions. Our demonstration scenarios start with basic Spannerlog solutions, then illustrate the power of \libname by easily combining Spannerlog and LLMs towards a real task (code generation), and finally analyze an existing \libname rewriting of an existing rule-based codebase where the code was simplified drastically (e.g., an order of magnitude fewer code lines).

\section{The Spannerlog Framework}

\libname is a library for expressing document spanners. In particular, it provides a concrete implementation of Spannerlog~\cite{nahshonIncorporatingInformationExtraction2016}, which is a variant of Datalog that operates on strings and spans in addition to ordinary relational data. In this section, we review the basic concepts of spanners and Spannerlog.

\def\spantriple#1#2#3{\langle#1,#2,#3\rangle}

Assume that we are given a set $D$ of \emph{text documents} (i.e., strings) and let  $\Sigma$ be the alphabet of all symbols used in these documents. 
A \emph{span} is a triple of the form $\spantriple{d}{i}{j}$ where $d \in D$ and $i$ and $j$
are naturals numbers such that $1 \leq i \leq j \leq |d|+1$, where $|d|$ is the length of $d$. The span $\spantriple{d}{i}{j}$ represents the position of the substring 
$d_i,\dots,d_{j-1}$ of $d$.

\def\insch{\mathit{in}}
\def\outsch{\mathit{out}}

By a \emph{schema} we mean here a sequence of types, where each type is either ``str'' (string) or ``span.''
An \emph{IE function} $f$ is a stateless function that takes a tuple over a fixed schema $\insch(f)$ and returns a relation over a fixed schema $\outsch(f)$ derived from the input tuple. For example, the standard NLP tool part-of-speech (POS) can be associated with an IE function $\mbox{POS}$ with $\insch(\mbox{POS})=(\mbox{str})$ and $\outsch(\mbox{POS})=(\mbox{span},\mbox{str})$. This IE function takes as input a string and returns the spans of every word over the input string and its part-of-speech tag as a string. IE functions that are widely studied in this context are regular expressions (regex formulas). If $\alpha$ is a regex formula, then $\mbox{rgx}_{\alpha}$ is a IE function with $\insch(\mbox{rgx}_{\alpha})=(\mbox{str})$ and $\outsch(\mbox{rgx}_{\alpha})=(s_1,\ldots,s_n)$, where $s_i=\mbox{span}$ and $n$ is the number of capture groups in the $\alpha$. Given an input document $d$, $\mbox{rgx}_{\alpha}(d)$ returns all matches of $\alpha$ in $d$ where each match $(\mbox{span}_1,\ldots,\mbox{span}_n)$ is ordered by the order of the first appearance of each capture variable in $\alpha$. 


To illustrate, given $\alpha=\mbox{x\{a+\}c+y\{b+\}}$ and an input string $d=\textsf{acb aacccbbb}$, $\mbox{rgx}_{\alpha}(d)$ will return the tuples  $(\spantriple{d}{0}{1},\spantriple{d}{2}{3})$ and $(\spantriple{d}{4}{6},\spantriple{d}{9}{12})$ mapping to $(\textsf{a},\textsf{b})$ and $(\textsf{aa},\textsf{bbb})$, respectively.
IE functions can be extended to handle other primitives (e.g., numbers), but we restrict to strings and spans for simplicity.


Assume a relational database schema.
A \emph{Datalog} program is collection of \emph{rules} of the form
$$p(x_1, \ldots, x_n) \leftarrow q_1(\tau_{1,1}, \ldots, \tau_{m_1,1}), \ldots, q_k(\tau_{1,k}, \ldots, \tau_{m_k,k})$$ where $p, q_1, \ldots, q_k$ are predicate symbols, each $x_i$ is a variable, and each $\tau_{i,j}$ can be either a variable or a constant. Rules can share the same left hand side predicate and can contain recursion. 

Spannerlog~\cite{nahshonIncorporatingInformationExtraction2016} is essentially Datalog over relations on strings and spans, extended with \emph{IE clauses} that refer to an execution of an IE function as a means of deriving facts using rules. 

\def\varn#1{\mathsf{#1}}
\def\rel#1{\mathsf{#1}}
\def\ief#1{\mathsf{#1}}
\def\scomma{\,,\,}
\def\maps{\mapsto}

More precisely, an \emph{IE predicate} is an expression of the form $f(x_1,\ldots,x_n)\maps(y_1,\ldots,y_m)$,
where $f$ is an IE function with an $n$-ary input schema and an $m$-ary output schema. Semantically, this predicate binds to $y_1,...,y_m$ all facts that can be derived from an application of $f$ over the values bound to $x_1,...,x_n$. In this work, we use the $\maps$ notation to make the distinction between the input and the output of an IE clause more clear.
A Spannerlog program is a set of Spannerlog rules, where a Spannerlog \emph{rule}
is defined similarly to a Datalog rule, except that rule bodies can include IE atoms. For illustration, consider the following rule.
\begin{align*}
\mathsf{
\rel{R}(\varn{usr},\varn{dom}) \gets
\rel{Texts}(\varn{d},\varn{t})
\scomma
\ief{rgx}_\alpha(\varn{t})\maps(\varn{usr},\varn{dom})
}
\end{align*}
Here, $\rel{Texts}$  is a binary relation that contains document texts ($\varn{t}$) and their dates of publication ($\varn{d}$). $\alpha$ is a regex formula that extracts the user and domain of email addresses found in the given text $\varn{t}$.

This extension of Datalog rules allows us to elegantly compose multiple IE functions as part of a rule in a readable way. For example, the following rule refers to the IE functions $\ief{rgx}_\alpha$ and $\ief{foo}$. It applies the regex extraction on the document produced by $\ief{foo}$.
\begin{align*}
\mathsf{
T(z,v,w)\gets 
    Texts(d,t)\scomma
    \ief{foo}(d,t)\maps(z)\scomma
    \ief{rgx}_\alpha(z)\maps(w,v)}
\end{align*}

\section{Main Library Components}
In this section, we introduce the main components of \libname. We first discuss the basic implementation of Spannerlog (\Cref{sec:spannerlog)}), then embedding of Spannerlog in Python (\Cref{sec:embed}), and then the incorporation of IE (callback) functions in Spannerlog (\Cref{sec:iefunctions}). 

\subsection{Spannerlog Implementation}\label{sec:spannerlog)}

The bedrock of \libname is a Python implementation of the Spannerlog language~\cite{nahshonIncorporatingInformationExtraction2016} which allows us to programmatically invoke the Spannerlog runtime from Python,  and vice versa.

To build our Spannerlog implementation, we extended the \emph{naive bottom up} evaluation method~\cite{DBLP:journals/ftdb/GreenHLZ13} to include evaluation of IE clauses. Since Spannerlog requires a more intricate definition of rule safety, which in turn determines IE function execution order within a rule, we also implemented a semantic safety checker according to the safety definitions in \cite{nahshonIncorporatingInformationExtraction2016}. 

Additionally, we added aggregation functions to \libname, and we will later illustrate their importance. For example, the next rule
concatenates all extracted $y$ in the same document $x$, in lexicographic order.
$$
\mathsf{
\rel{R}(t,\ief{lex\_concat}(\mathsf{str}(y)))\gets 
\rel{Texts}(d,t)\scomma
\ief{rgx}_\alpha(t)\maps(y)
}
$$
Note that the syntax and semantics are similar to previous Datalog formalisms for aggregation (e.g.,\citeN{DBLP:conf/icde/ShkapskyYZ15}).

\subsection{Embedding Spannerlog in Python}\label{sec:embed}

To be useful in the setting of a Python program, we embedded our Spannerlog engine inside the Python runtime, via IPython's \emph{Magic} system. The Magic system allows us to write and register Spannerlog rules directly inside a Jupyter Notebook Cell. 

\libname uses the markers \textsf{\%\%python} and \textsf{\%\%log} to separate between Python code and Spannerlog rules. The session object facilitates communication between the Python and Spannerlog runtimes. As an example, consider the following code.

\definecolor{darkblue}{rgb}{0, 0, 0.8}
\definecolor{darkgreen}{rgb}{0, 0.5, 0}
\definecolor{darkgrey}{rgb}{0.4, 0.4, 0.4}

\def\magic#1{{\color{darkgrey}\%\%#1}}

\def\code#1{
\smallskip
{\parbox{3.0in}{
\hrule\smallskip
{\sf\color{black}
#1}
\smallskip\hrule
}}
\bigskip
}

\code{
\magic{python}\\
from rgxlog import session\\
df = pd.Dataframe(\dots columns =[`Date',`Text']) \\
session.import(df,name=`Texts')\\
\magic{log}\\
R(usr,dom) $\gets$ Texts(d,t)\,,\, rgx\_alpha(t) $\maps$ (usr,dom)\\
\magic{python}\\
Out = session.export(`?R(usr,``gmail'')')
}

This code constructs the \textsf{Texts} relation by taking a DataFrame and importing it to our engine, applies the Spannerlog rule to construct the relation \textsf{R}, and then populates the DataFrame \textsf{Out} by querying \textsf{R} (asking for users whose email domain is ``gmail'').

\subsection{Embedding Python in Spannerlog}\label{sec:iefunctions}

While Spannerlog is useful as an end-to-end text analysis engine, there are tasks that require calling external functionality, often provided by machine-learning-based Natural Language Processing (NLP) models. These models are often accessible as software programmed or wrapped in Python. 
\libname enables the user to empower Spannerlog with user-defined functions and integrate them within as novel callbacks to IE functions. Any stateless Python function can be added as an IE callback via the session object, as long as the function accepts a relation as input and returns a relation (and more generally sequence of relations) as output. Following is an example.

\def\ind{\indent\hspace{2em}}

\code{
\magic{python}\\
def foo(x:str,y:str):\\
\quad \dots \\
session.register(foo, input=[str,str], output=[span])\\
\magic{log}\\
T(z,v,w) $\gets$
    R(x,y),
    S(`bob',x),
    $\ief{foo}$(x,y) $\maps$ (z), \\
    \ind $\ief{rgx\_alpha}$(z) $\maps$ (w,v) \\
\magic{python}\\
session.export(`?T(z,w,``gmail'')')
}

\section{Library Illustration}
We next illustrate the utility of \libname by describing an example of augmenting the context of an LLM as well as a concrete use case of rewriting an existing rule based NLP pipeline using \libname.

\subsection{Example: Code Documentation Task} \label{sec:copilot}

We first illustrate how \libname 
can help build LLM pipelines by concisely expressing programmer intent and, in particular, avoiding orchestration boilerplate and data management.

Consider the task of developing a code completion program focused on suggesting documentation for the function currently containing the user cursor in a code editor, similar to Github's Copilot.\footnote{\url{https://github.com/features/copilot}}
One of the challenges of performing code completions is that the entire code base is too big to fit into the context window. Copilot's approach to solving this issue is to take the last $k$ files that were accessed in the code editor (including the current file where the cursor is pointing to) and feed them all as context to the LLM model. This approach has two main weaknesses: \emph{(1)} there would mostly be a lot of irrelevant context in some of these files, \emph{and (2)} there could be very relevant instances of interacting with the function among the files that have not been touched recently.

We will illustrate the simplicity and elegance of implementing a more sophisticated context retrieval approach. We would like to provide the LLM with the following context that consists of two components:
the code of the current function, and the code of all functions that call the current function.
This task, like most LLM tasks and specifically Code LLM tasks, requires integration of basic LLM core functionalities (like a chat model API), can leverage basic human understanding of the text, and is unique in the orchestration of the specific solution while using very similar tooling as other solutions in the same field. For the illustration, let us assume we have the following relations:
\begin{itemize}
\item A binary relation \textsf{Files(name, content)} that holds all file names in the code project and their content as strings;
\item A unary relation \textsf{Cursor(pos)} that holds a single span over the files with the position of our cursor. 
\end{itemize}
Moreover, let us also assume some additional IE functions, which can be constructed by very thin wrappers around established libraries: the IE function \textsf{LLM}  calls an LLM with a prompt and returns the answer as a string, and the IE function \textsf{AST} takes an XPath pattern and a document and returns all spans of code that fits the AST pattern. For example, the pattern \textsf{.*.(FuncDecl|ClassDecl)}  returns all function and class definitions, nested in the AST.
We also assume as IE functions some standard operations such as string concatenation, span containment, and a printf-like formatting.


Using these generic primitives, it is quite simple and elegant to convey our pipeline's logic via \libname. We show here only parts of the code for brevity, the full \libname implementation will be presented as part of the demonstration (see \Cref{scenarios}).

\code{
\magic{python}\\
def llm(q):\\
    \quad \dots \\
session.register(llm, input=[str], output=[str])\\
def ast(pattern,doc):\\
    \quad \dots \\
session.register(ast,input=[str,str],output=[span])\\
\magic{log}\\
scope\_of(pos,s) $\gets$
    Files(name,c) \scomma \\
    \ind AST(`.*.(FuncDecl|ClassDecl)',c)
    $\maps$(s) \scomma
        contains(pos,s)\\
document(p,a)$\gets$
    scope\_of(p,s)\scomma
    get\_name(p,name)\scomma\\
    \ind mentions(name,m)\scomma
    format(prompt,m,c)$\maps$(q)\scomma\\
    \ind LLM(q)$\maps$ (a)\\
\magic{python}\\
session.export(``?document(pos,a)'')
}

\subsection{Case Study} \label{sec:nlp-usecase}

To study the ability of \libname to simplify text analysis tasks on a bigger scale, we took the source code for an existing rule-based NLP pipeline and rewrote it with \libname. The original pipeline~\cite{chapmanNaturalLanguageProcessing2020} deals with analyzing medical text documents to classify patients based on their COVID-19 risk. The pipeline consists of 4335 lines of Python code and is based on the open-source NLP library spacy~\cite{honnibal2020spacy}. 
The full rewriting process and the refactored code are available online.\footnote{ \url{https://deanlight.github.io/spanner_workbench/covid-nlp/covid_pipeline.html}}

\Cref{tab:nlp-code-comparison} shows a summary of the code-line comparison between the implementations. For a fair comparison, we kept library usage as close as possible to the original code. As can be seen, under 10\% of the original code was either core computations or interfaces, which remained as Python code (203 lines). Over 90\% of the code was either data management or pipeline orchestration (including specific constants supplied to the pipelines in different steps). Hence, we were able to reduce around $4000$ lines of code into fewer than $400$ lines of either Spannelog rules or tabular data. 

\begin{table}
\small
    \caption{Code comparison between the original implementation  \cite{chapmanNaturalLanguageProcessing2020} and the rewritten \libname implementation.}
    \centering
    \begin{tabularx}{0.48\textwidth}{|X|c|c|} \hline 
         \textbf{Code Type}&  \textbf{Original Code}&  \textbf{\libname}\\ \hline 
         Native Python&  4335&  110\\ \hline 
         Python IE Functions&  &  93\\ \hline 
 Spannerlog Code& &107\\ \hline 
 Code As Data (txt/csv files)& &286\\ \hline 
 \textbf{Total Python}& \textbf{4335}&\textbf{203}\\ \hline 
 \textbf{Total Declarative Code}& &\textbf{393}\\ \hline 
 \textbf{Total Lines}& \textbf{4335}&\textbf{596}\\ \hline
    \end{tabularx}

    \label{tab:nlp-code-comparison}
\end{table}

\section{Demo Scenarios} \label{scenarios}

All demonstration scenarios are interactive Jupyter notebooks that contain step by step walkthroughs of how to design and implement increasingly complex programs using \libname.

\medskip\noindent\textbf{Basic Task.}
In this scenario, we showcase the basic components of \libname, in particular defining and using IE functions and composing them via Spannerlog code. This scenario will include simple tasks such as finding identical sentences in a corpus of documents and incorporating LLMs like ChatGPT in a question answering pipelines. 




\medskip\noindent\textbf{End-to-End Task.}
In this scenario, we show a full working implementation of the code documentation task explained in \Cref{sec:copilot}.
We will use the \textsf{LLM} IE function from the previous scenario as well as wrap Python's AST library into an IE function $\mbox{AST}$. We will then show how we express our context augmentation pipeline's logic as Spannerlog rules and how we query our rules to receive the output of the LLM and export it to Python. We also analyze our code for readability and conciseness. This scenario demonstrates how complex data pipelines can easily be expressed using \libname.

\medskip\noindent\textbf{Extending \libname Code.}
While good code is clear and concise, it should also make modifications and extensions easy for the programmer.
In this scenario we demonstrate the ease with which \libname code can be extended.
To do so, we build upon the previous code documentation implementation and extend it to include: (1) augmentation of the LLM's context according to data sources that were not available to the LLM during training. (2) User feedback over previous executions of this task. These extensions will use two of the most widely used prompt augmentation techniques used in the industry today, namely Retrieval Augmented Generation (RAG)~\cite{lewis2021retrievalaugmented} and Few Shot Prompting~\cite{brown2020language}.




This scenario will not only demonstrate the ease of extension of \libname code but will also demonstrate that modern techniques from the field of NLP can be easily expressed and composed via \libname to generate sophisticated pipelines that result in a very simple code base.

\medskip\noindent\textbf{Real Code Base.}
In this scenario, we aim to quantify the benefits of integrating \libname into a code base from a software engineering perspective. Concretely, we will demonstrate how to refactor an existing code base using \libname, by means of the rule-based medical document classification pipeline of \Cref{sec:nlp-usecase}.  We will show a side-by-side comparison of the original version and the \libname rewriting. 
Using this comparison, we will explain and demonstrate best practices for 
decomposing existing code into \libname, namely: (1) How to identify the scope of and implement IE functions? (2) Which code should turn into declarative code, either as Spannerlog rules and queries or as data? (3) What parts of the code base should be left as external Python code? 

We will then analyze the size reduction ratio of different fragments of code (as seen in \Cref{tab:nlp-code-comparison}) as they are translated into \libname code. We will also discuss how both implementations differ in various code quality measures such as conceptual decomposition, separation of concerns, bug surface area, and so on. This scenario will demonstrate that the paradigm introduced by \libname is robust and powerful enough to significantly improve development even in real-world code bases.


\section{Discussion}
While rule-based IE systems have been shown to increase processing performance and have been argued to increase developer productivity \cite{chiticariu-etal-2018-systemt}, they have not seen wide adoption. We believe this stems from a lack of accessible tooling and developer education. To remedy the former, \libname (1) embeds itself within the python ecosystem, removing many of the barriers of writing IE functions and (2) leverages an extremely simple declarative language as its backbone. To remedy the latter, we discuss IE based programming from a software engineering perspective in our walkthroughs, using lines of code as a surrogate metric for ease of development. 
\libname does not yet put an emphasis on processing performance, and is therefore not comparable to existing enterprise IE systems in terms of processing efficiency. We are addressing this topic in ongoing work.




\begin{acks}
This work was funded by the Israel Science Foundation (ISF) under grant 768/19. The work of Dean Light and Benny Kimelfeld has been funded by the German Research Foundation (DFG) under grant KI 2348/1-1. S.~Vansummeren was supported by the Bijzonder Onderzoeksfonds (BOF) of Hasselt University under Grant No. BOF20ZAP02 as well as the Flanders AI research program.
\end{acks}


\bibliographystyle{ACM-Reference-Format}
\bibliography{sample}

\end{document}